\pdfoutput=1
\documentclass[11pt,a4paper]{article}
\usepackage{acl2017}
\usepackage{times}
\usepackage{latexsym}

\aclfinalcopy % Uncomment this line for the final submission
\def\aclpaperid{9} %  Enter the acl Paper ID here

\title{Toward Automated Early Sepsis Alerting: Identifying Infection Patients from Nursing Notes}

\author{Emilia Apostolova \\
  Language.ai \\
  Chicago, IL, USA\\
  {\tt emilia@language.ai} \\\And
  Tom Velez \\
  Vivace Health Solutions \\
  Cardiff, CA, USA \\
  {\tt tom.velez@cta.com} \\}

\date{}

\begin{document}
\maketitle
\begin{abstract}

Severe sepsis and septic shock are conditions that affect millions of patients and have close to 50\% mortality rate. Early identification of at-risk patients significantly improves outcomes. Electronic surveillance tools have been developed to monitor structured Electronic Medical Records and automatically recognize early signs of sepsis. However, many sepsis risk factors (e.g. symptoms and signs of infection) are often captured only in free text clinical notes. In this study, we developed a method for automatic monitoring of nursing notes for signs and symptoms of infection. We utilized a creative approach to automatically generate an annotated dataset. The dataset was used to create a Machine Learning model that achieved an F1-score ranging from 79 to 96\%. 
 
\end{abstract}

\section{Introduction}

Severe sepsis and septic shock are rapidly progressive, life-threatening conditions caused by complications from an infection. They are major healthcare problems that affect millions of patients globally each year \cite{kim2016sepsis}. The mortality rate for severe sepsis and septic shock is approaching 50\% \cite{nguyen2006severe}.

A key goal in critical care medicine is the early identification and treatment of infected patients with early stage sepsis. The most recent guidelines for the management of severe sepsis and septic shock include early recognition and management of these conditions as medical emergencies, immediate administration of resuscitative fluids, frequent reassessment, and empiric antibiotics as soon as possible following recognition \cite{dellinger2008surviving}.

% Timely treatment of sepsis and early identification of patients at risk of ARDS significantly optimize outcomes. The guidelines for the management of severe sepsis and septic shock include early recognition and management, continuous monitoring of vital signs, laboratory testing,  clinical evaluation, imaging, as well as the administration of empiric antibiotics as soon as possible.

Early recognition of infections that can lead to sepsis, severe sepsis and/or septic shock can be challenging for several reasons: 1) these conditions can quickly develop from any form of common infections (bacterial, viral or fungal) and can be localized or generalized; 2) culture-dependent diagnosis of infection is commonly slow and prior use of antibiotics may make cultures falsely negative \cite{vincent2016clinical}; 3) systemic inflammatory response syndrome, traditionally associated with sepsis, may be the result of other noninfectious disease processes \cite{bone1992accp}. Consequently, clinicians frequently rely on a myriad of non-specific symptoms of infections and physiological signs of systemic inflammatory response for rapid diagnosis. Each hour of delay in the administration of recommended therapy is associated with a linear increase in the risk of mortality rate \cite{kumar2006duration,han2003early}, driving the need for automation of early sepsis recognition.

In response to this need, electronic surveillance tools have been developed to monitor for the arrival of new patient electronic medical record (EMR) data, automatically “recognize” early signs of sepsis risk in specific patients, and trigger alerts to clinicians to help guide timely, effective interventions \cite{herasevich2011limiting,azzam2009validation,koenig2011performance}. 
The automated decision logic used in many existing sepsis screening tools, for example \cite{nguyen2014automated,hooper2012randomized,nelson2011prospective}, relies on consensus criteria-based rules. 

Structured EMR data, such as diagnostic codes, vital signs and orders for tests, imaging and medications, can be a reliable source of sepsis criteria. However, many sepsis risk factors (e.g. symptoms and signs of infection) are routinely captured solely in free text clinical notes. The aim of this study is to develop a system for the detection of signs and symptoms of infection from free-text nursing notes. The output of the system is later used, in conjunction with available structured data, as an input to an electronic surveillance tool for an early detection of sepsis.

\section{Task Definition and Dataset}

Depending on the infection source and the specifics of the patient history, signs and symptoms of infection can vary widely. In addition, similar symptoms can be stated using a number of synonymous expressions, complicated by the presence of abbreviations, variant spellings and misspellings. Table \ref{table1} lists nursing note snippets indicating a possible presence of infection with various degrees of certainty. For example, line items 1, 8, and 12 indicate increased temperature; line items 1, 3, 6, 7, and 13 indicate the use of infection-treating antibiotics; line items 4, 7, 9, 11, and 13 mention specific infectious diseases. A number of examples mention additional infection symptoms or infection detecting tests.

\begin{table}[h]
	\small
\begin{center}
\begin{tabular}{|l|l|}
	\hline
1 & Afebrile on antibiotics.\\
\hline
2 & Very large copious amount of secretions,sputum\\
\hline
3 & ... medicated with iv cefazolin dose 2 of 3\\
\hline
4 & UA positive for UTI.\\
\hline
5 & Blood culture pos for gram neg  organisms.\\
\hline
6 & Elevated WBC count, on clindamycin IV.\\
\hline
7 & ... continues on clindamycin(D6), pen-G(D5) \\
& and doxycycline(D4) for LLL pneumonia\\
\hline
8 & 84 year old male with h/o throat cancer who \\
& presented on [DATE] to [LOCATION] with fever, \\
& diffuse rash, renal failure and altered mental status.\\
\hline
9 & Pt had a positive sputum specs for GBS and GPC, \\
& and he has HSV on lips.\\
\hline
10 & ... blood, urine and sputum culture sent today ...\\
\hline
11 & PEG tube to gravity, episodes of vomitting on day\\
& shift,NPO. Contact precautions MRSA.\\
\hline
12 & Pt had temp spike of 102.4\\
\hline
13 & Levaquin started for pnuemonia.\\
\hline
14 & Had infected knee prosthesis which led to wash out \\
& of joint yesterday.\\
\hline
\end{tabular}
\end{center}
\caption{\small MIMIC-III nursing note snippets indicating a presence of infection with various degrees of certainty. Abbreviations and misspellings are preserved to demonstrate the task challenges.}
\label{table1}
\end{table}

In the literature of identifying patient phenotype cohorts using electronic health records, most studies map textual elements to standard vocabularies, such as the Unified Medical Language System (UMLS) \cite{shivade2014review}. The standard vocabulary concepts are later used in rule-based, and, in some cases, Machine Learning (ML) approaches to identify patient cohorts.

In the context of identifying infection from clinical notes, however, such an approach poses a number of challenges. Symptoms can vary widely depending on the source of infection, for example, \textit{redness, sputum, swelling, pus, phlegm, vomiting, increased white blood cell count}, etc. The same symptom can also be expressed in a large number of ways, for example, \textit{afebrile, temp spike of 102.4, fever}, etc. There is a large number of conditions indicating infections, for example \textit{UTI, strep throat, hepatitis, HIV/AIDS}, etc. In addition, abbreviations and misspellings are quite common in the context of ICU care, for example, \textit{pneumonia, PNA, pnuemonia, pneu}, etc.

Due to their nature, the dataset and task are better suited for ML approaches that are not relying on standard vocabularies or a structured set of features. As with most ML tasks in the clinical domain, the challenge in this approach is obtaining a sufficient amount of training data \cite{chapman2011overcoming}. 

To address these challenges, we utilized the MIMIC-III dataset \cite{johnson2016mimic} and developed a creative solution to automatically generate training data as described in section \ref{subsection:rules}. MIMIC (Medical Information Mart for Intensive Care) is a large, freely-available database comprising deidentified health-related data associated with over 40,000 patients who stayed in critical care units of the Beth Israel Deaconess Medical Center between 2001 and 2012. The dataset contains over 2 million free-text clinical notes. We focused only on nursing notes for adult patients, and our dataset consists of a total of 634,369 nursing notes.

\section{Rule-based Training Dataset Creation}
\label{subsection:rules}

To obtain a sizable training dataset we explored the use of available MIMIC-III structured data, such as test orders and results, prescribed medications, and diagnosis codes. However, this approach did not translate to accurately identifying nursing notes suggesting infection for a number of reasons\footnote{Challenges include missing or incorrect data, discontinuous or disordered EMR data entry timestamps, etc.}. Instead, we utilized a simple heuristic. We observed that whenever there is an existing infection or a suspicion of infection, the nursing notes describe the fact that the patient is taking or is prescribed infection-treating antibiotics. Thus, identifying nursing notes describing the use of antibiotics will, in most cases, also identify nursing notes describing signs and symptoms of infection. 

To identify positive mentions of administered antibiotics, we used a list of the 60 most commonly administered infection-treating antibiotics in the MIMIC dataset \cite{Misquitta2013}. This initial list was then extended to include additional antibiotic names, brands, abbreviations, spelling variations, and common misspellings. We semi-automated this laborious task by utilizing \textit{word embeddings} \cite{mikolov2013distributed}. Word embeddings were generated utilizing all available MIMIC-III nursing notes\footnote{We used vector size 200, window size 7, and continuous bag-of-words model.}. The initial set of antibiotics was then extended using the closest \textit{word embeddings} in terms of cosine distance.  For example, the closest words to the antibiotic \textit{amoxicillin} are \textit{amox, amoxacillin, amoxycillin, cefixime, suprax, amoxcillin, amoxicilin}. As shown, this includes misspellings, abbreviations, similar drugs and brand names. The extended list was then manually reviewed. The final infection-treating antibiotic list consists of 402 unambiguous expressions indicating antibiotics. 

Antibiotics, however, are sometimes negated and are often mentioned in the context of allergies (e.g. \textit{allergic to penicillin}). To distinguish between affirmed, negated, and speculated mentions of administered antibiotics, we also developed a set of rules in the form of keyword triggers. Similarly to the NegEx algorithm \cite{chapman2001simple}, we identified phrases that indicate uncertain or negated mentions of antibiotics that precede or follow a list of hand-crafted expression at the sentence and clause levels. Word embeddings were again used to extend the list of triggers with synonyms, spelling variations, abbreviations, and misspellings. For example, the words \textit{allergic, anaphylaxis, anaphalaxis, allerg, }and \textit{anaphylaxsis} are all used as triggers indicating the negation of an antibiotic use. The full list of keywords indicating antibiotics, negation/speculation triggers and conjunctions is available online\footnote{https://github.com/ema-/antibiotic-dictionary}.

The described approach identified 186,158 nursing notes suggesting the unambiguous presence of infection (29\%) and 3,262 notes suggesting possible infection. The remaining 448,211 notes (70\%) were considered to comprise our negative dataset, i.e. not suggesting infection.
\section{Machine Learning Results}
We modeled the task as a binary classification of free-form clinical notes. It has been shown that Support Vector Machines \cite{cortes1995support} achieve superior results in most text classifications tasks and were selected as a sensible first choice. The individual nursing notes were represented as a bag-of-words (1-grams). The tokens were all converted to lower case and non-alphanumeric characters were discarded. Tokens that are present in more than 60\% of all samples or less than 6 times were also discarded. The tokens were weighted using the tf-idf scheme \cite{salton1986introduction}. We trained the model using linear kernel SVMs\footnote{We used the LibSVM library with both the cost and gamma parameters set to 2 (obtained via grid-search parameter estimation).} \cite{CC01a}. We set the positive class weight to 2 to address the unbalanced dataset. 70\% of the automatically generated dataset was used for training and the remaining 30\% for testing. This resulted in a precision of 93.12 and a recall of 99.04 as shown in Table \ref{table3}.

\begin{table}[h]
	\small
\begin{center}
\begin{tabular}{l|l|l|l}
     & \textbf{Precision}  & \textbf{Recall} & \textbf{F1-score}\\ \hline
	SVM\textsuperscript{auto} &  93.12 & 99.04 & 95.99  \\
	SVM\textsuperscript{gold} &  92.10 & 68.46 & 78.53 \\
\end{tabular}
\end{center}
\caption{\small Classification Results. SVM\textsuperscript{auto}=Results from applying the SVM model on an automatically generated test set of 190,000 nursing notes; SVM\textsuperscript{gold}=Results from applying the SVM model on a manually reviewed dataset of 200 nursing notes.}
\label{table3}
\end{table}

As the training dataset was automatically created, the above results do not truly reflect the model performance. To evaluate the model on the ground truth, a qualified professional manually reviewed 200 randomly selected nursing notes. These results are also shown in Table \ref{table3}. While the model precision remained high (92.10), the recall dropped significantly to 68.46. 

The drop in recall can be partially attributed to the manner in which the testing data was created. Nursing notes describing signs of infection but failing to mention the use of antibiotics were considered (incorrectly) negative examples. However, an error analysis revealed that the majority of the false negatives (contributing to the low recall) were actually all indicating low level of suspicion of infection. For example, the human annotator considered the following snippets sufficient to indicate a possible infection \textit{
afebrile, bld cx's sent; monitor temp, wbc's, await stool cx results; lungs coarse, thick yellow secretions suctioned from ett; awaiting results of CT, malignancy vs pneumonia}. In all cases, the note expresses only a suspicion for infection, pending further tests.

We further attempted to improve the system performance by utilizing Paragraph Vectors \cite{le2014distributed}. Unsupervised algorithms have been used to represent variable pieces of texts such as paragraphs and documents as fixed-length feature representations (Paragraph Vectors). Studies have shown that Paragraph Vectors outperform bag-of-words models on some text classification tasks. We used the text from all nursing notes to create Paragraph Vectors. We generated document embeddings using a distributed memory model and distributed bag-of-words model, each of size 300 with a window size of 7. Combining the vectors of the distributed memory model and the distributed bag-of-words model, we represented each document as a vector of size 600. The paragraph vectors of the training instances were then fed to a logistic regression, K-nearest neighbors, and an SVM classifier. Results significantly under-performed the SVM bag-of-words model and we were able to achieve a maximum precision and recall of 63\% and 77\% respectively.

\section{Related Work}

A review of approaches to identifying patient phenotype cohorts using EMR data \cite{shivade2014review} describes a number of studies using clinical notes, most often in combination with additional structured information, such as diagnosis codes. The study asserts that clinical notes are often the only source of information from which to infer important phenotypic characteristics.

Demner-Fushman et al. \shortcite{demner2009can} note that clinical events monitoring is one of the most common and essential tasks of Clinical Decision Support systems. The task is in many respects similar to the task of identifying patient phenotype cohorts and it has been observed that free text clinical notes are again the best source of information. For example, Murff et al. \shortcite{murff2003detecting} found the electronic discharge summaries to be an excellent source for detecting adverse events. They also note that simple keywords and triggers are not sufficient to detect such events.

In the context of identifying infection from clinical text, most studies map textual elements to standard vocabularies, such as UMLS. For example, Matheny et al. \shortcite{matheny2012detection} develop a system for detecting infectious symptoms from emergency department and primary care clinical documentation, utilizing keywords and SNOMED-CT concepts. Bejan et al. \shortcite{bejan2012pneumonia} describe a system for pneumonia identification from narrative reports using n-grams and UMLS concepts. Similarly, Elkin et al. \shortcite{elkin2008nlp} encoded radiology reports using SNOMED-CT concepts and developed a set of rules to identify pneumonia cases. 

Horng et al. \shortcite{horng2017creating} develop an automated trigger for sepsis clinical decision support at emergency department triage. They utilize machine learning and establish that free text drastically improves the discriminatory ability of identifying infection (increase in AUC from 0.67 to 0.86). Arnold et al. \shortcite{arnold2014comprehensive} develop an EHR screening tool to identify sepsis patients. They utilize NLP applied to clinical documentation, providing greater clinical context than laboratory and vital sign screening alone. DeLisle et al. \shortcite{delisle2010combining} used a combination of structured EMR parameters and text analysis to detect acute respiratory infections. Murff et al. \shortcite{murff2011automated} develop a natural language processing search approach to identify postoperative surgical complications within a comprehensive electronic medical record.

Halpern et al. \shortcite{halpern2014using} describe a system for learning to estimate and predict clinical state variables without labeled data. Similar to our approach, they use a combination of domain expertise and vast amounts of unlabeled data, without requiring labor-intensive manual labeling. In their system, an expert encodes a certain amount of domain knowledge (identifying anchor variables) which is later used to train classifiers. Elkan and Noto \shortcite{elkan2008learning} show that a classifier trained on positive and unlabeled examples predicts probabilities that differ by only a constant factor from the true conditional probabilities of being positive.

\section{Discussion}

We presented an approach to identifying nursing notes describing the suspicion or presence of an infection. We utilized the MIMIC-III dataset and a creative approach to obtain an ample amount of annotated data. We then applied ML methods to the task and achieved performance sufficient for practical applications. The ultimate goal of this study is to utilize free-text notes, in combination with structured EMR data, to build an automated surveillance system for early detection of patients at risk of sepsis.

% \section*{Acknowledgments}
%
% The acknowledgments should go immediately before the references.  Do
% not number the acknowledgments section. Do not include this section
% when submitting your paper for review.

% include your own bib file like this:
%\bibliographystyle{acl}
%\bibliography{acl2017}
\bibliography{acl2017}
\bibliographystyle{acl_natbib}

\end{document}